\documentclass[twocolumn,preprintnumbers,amsmath,amssymb]{revtex4}
\usepackage{graphicx}
\usepackage{graphicx,ulem,amsmath,amssymb,amsfonts,xcolor}

\usepackage{bm}			
\bibpunct{}{}{,}{s}{}{} 

\newcommand\beq{\begin{equation}}
\newcommand\eeq{\end{equation}}

\begin{document}
	\title{Collimated hot electron generation from sub-wavelength grating target irradiated by a femtosecond laser pulse of relativistic intensity}
	
	\author{Kamalesh Jana$^1$}
	\author{Amit D. Lad$^1$, Guo-Bo Zhang$^2,^3$, Bo-Yuan Li$^2$, V. Rakesh Kumar$^1$, Moniruzzaman Shaikh$^1$, Yash M. Ved$^1$}
	\author{Min Chen$^2$}  \email{minchen@sjtu.edu.cn}
	\author{G. Ravindra Kumar$^1$}  \email{grk@tifr.res.in}
	
	\affiliation{$^1$Tata Institute of Fundamental Research, Dr. Homi Bhabha Road, Colaba, Mumbai-400005, India}
	
	\affiliation{$^2$Key Laboratory for Laser Plasmas (Ministry of Education), School of Physics and Astronomy, Shanghai Jiao Tong University, Shanghai 200240, China and Collaborative Innovation Center of IFSA, Shanghai Jiao Tong University, Shanghai 200240, China}
	
	\affiliation{$^3$Department of Nuclear Science and Technology, National University of Defense Technology, Changsha 410073, China}

	\begin{abstract}
		{We investigate the production of hot electrons from the interaction of relativistically-intense ($I> 10^{18} W/cm^{2}$) ultra-short (25 fs) laser pulses with sub-wavelength grating target. We measure the hot electron angular distribution and energy spectra for grating target and compare them with those from a planar mirror target. We observe that hot electrons are emitted in a collimated beam along the specular direction of the grating target. From the measured electron energy spectra we see electron temperature for grating is higher than the mirror, suggesting a higher electron yield and hence a stronger coupling with the laser. We performed numerical simulations which are in good agreement with experimental results, offer insights into the acceleration mechanism by resulting electric and magnetic fields. Such collimated fast electron beams have a wide range of applications in applied and fundamental science.}
	\end{abstract}

	\maketitle
	
	\section{Introduction}

	Ultrashort pulse high-intensity laser produced plasma is a novel source of relativistic electrons.\cite{Kruerbook,WilkIEEE1997,KawRMPP2017} These highly energetic particles of femtosecond duration underpin numerous applications such as ultrafast time resolved electron diffraction,\cite{SiwickScience2003} laboratory astrophysics,\cite{Drakebook} laser fusion,\cite{LeiPRL2006} medical therapy, x-ray sources,\cite{ChenPRL2008} femtochemistry\cite{ZewailScience1988} etc.\cite{KawRMPP2017} All these applications would greatly benefit from highly efficient sources of relativistic electrons. It is well known that these relativistic electrons are generated by high intensity laser absorption in the hot, dense plasma via various collisionless mechanisms.\cite{Kruerbook,WilkIEEE1997,KawRMPP2017} Many efforts\cite{RajeevPRL2003,MondalPRB20011,KahalyPRL2008,GahnAPL1998} have therefore been devoted to enhance hot electron generation via improved laser coupling to the plasma. One of the most efficient methods uses targets with rough surfaces composed with nanoparticles, nanowires and gratings \cite{RajeevPRL2003,MondalPRB20011,KahalyPRL2008}. Nanostructuring the target surface can lead to an enhancement in the coupling efficiency as physical mechanisms like surface plasmon excitation \cite{Maierbook} or lightning rod effect\cite{BoydPRB1984} locally intensify the incident electric field. Previous works with sub-wavelength grating targets with a controllable periodic structure have shown considerable enhancement of laser absorption as well as enhancement of hot electrons and other secondary emissions.\cite{KahalyPRL2008,HuPOP2010,BagchiPOP2012,WangPOP2008,CerchezPRL2013,YeungNJP2013} Early experiments were performed at moderate laser intensities.\cite{KahalyPRL2008,HuPOP2010,BagchiPOP2012} One of the laser parameters that controls the interaction at high intensities is the intensity contrast, considered crucial for limiting the pre-pulse and preserving the nano-structures on the target surface for the main pulse. Recent improvements in laser intensity contrast have propelled studies on nano-wire and grating targets at relativistic intensities. \cite{FedeliPRL2016,LadSciRep2022,MacchiPOP2018} In addition to large fluxes, many practical applications demand low divergence electron beams though often, the high-intensity laser-driven dense plasmas generate energetic electrons with broad divergence. 
	
	In this paper, we report generation of collimated relativistic electron beams from a grating target irradiated by a 25 fs relativistic intensity laser pulse. Measurements of angular distributions of hot electrons show that collimated hot electrons are emitted along the specular reflection direction of the grating target. We also present electron energy spectra and compare all the results with those from an optically polished planar gold mirror target. Efficient generation of higher energy, better collimated electrons from the grating target is demonstrated, indicating enhanced laser coupling to the gratings and beam steering by the resulting electric and magnetic fields at the grating surface. To understand the results, we performed particle-in-cell (PIC) simulations, which are in good agreement with the experimental results. \\
	
		\begin{figure}
		
		\includegraphics[width=1.0\columnwidth]{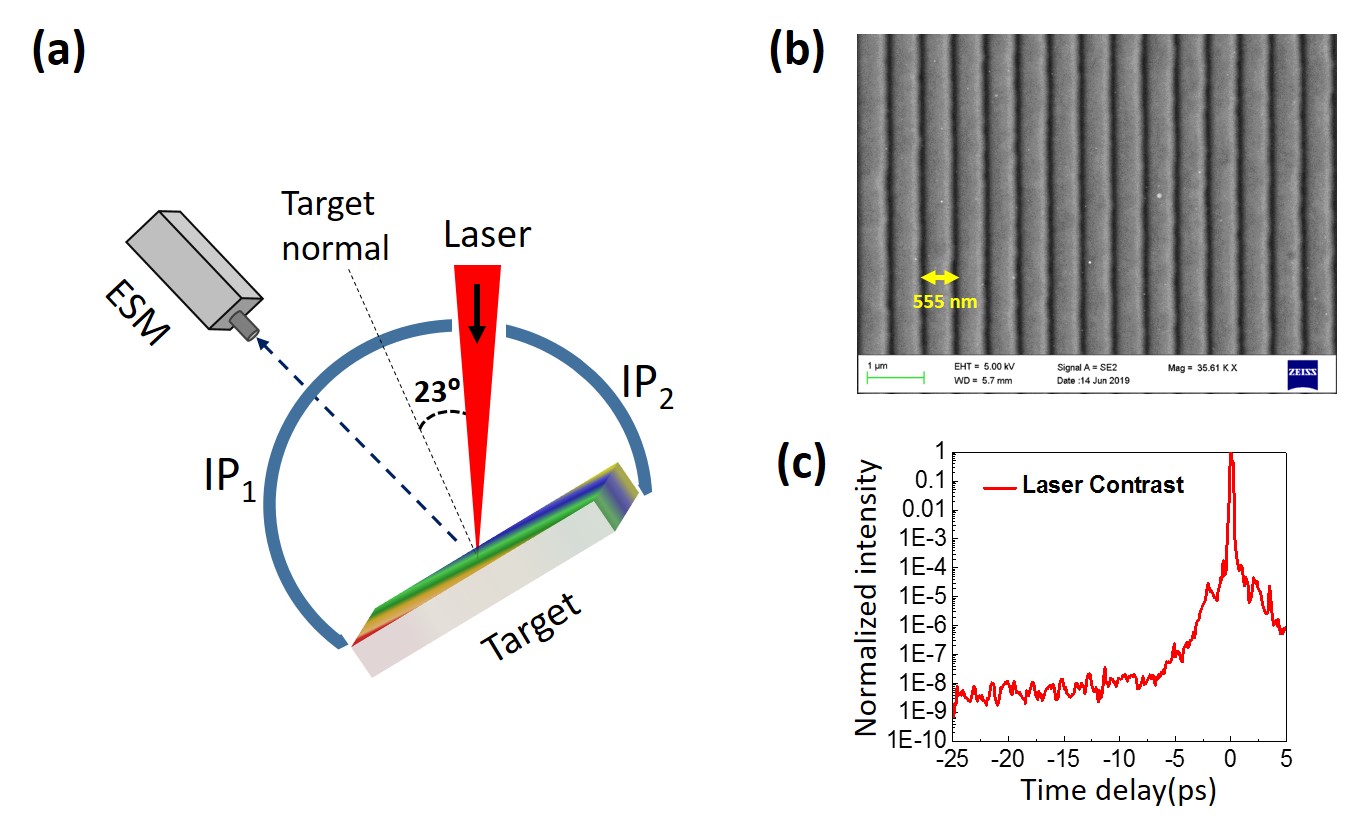}
		\caption{(a) Schematic of the experimental setup for measuring angular distributions and energy spectra of hot electrons; IP: Image Plate, ESM: Electron Spectrometer, Target: Gold coated grating or plane mirror. (b) Scanning electron microscope (SEM) image of the grating target ( Period: 555 nm ). (c) Laser intensity contrast measured by a third-order cross-correlator (SEQUOIA).}
		\label{figure1}

	\end{figure}

	\section{Experiment}
	
	The experiment [Fig. 1(a)] was carried out at the Tata Institute of Fundamental Research (TIFR), Mumbai using a 100 TW CPA (chirped pulse amplification) based laser systems (10 Hz). A p-polarized (800nm, 25 fs) laser pulse was focused by an f/2.5 off-axis parabolic mirror to a focal spot of $\sim$ 14 $\mu$m (FWHM). The intensity contrast of the incident laser pulse [measured by a third-order cross-correlator (SEQUOIA)] was  $\sim$ 10$^{-9}$ at 25 ps [Fig. 1(c)]. Triangular blazed grating with gold coating on a glass substrate (Jobin Yvon, 1800 lines/mm groove density, 555 nm period, 158 nm groove depth,  17.45$^{\circ}$ blaze angle) and polished (surface finish at $\lambda$/10) gold coated glass (gold mirror) were used as targets in this experiment. The thickness of the gold coating layer in both cases was many times the optical skin depth of $\sim$ 22 nm at 800 nm. The target was mounted on a four axis (X,Y,Z, $\theta$) motorized high-precision translation stage so that every laser pulse interacted with a fresh position on the target. The resonance angle (where the absorption is maximum) of the grating target was measured to be $\sim$ 23$^{\circ}$. It was fixed as the angle of incidence of the interacting laser pulse for all the measurements. The angular distribution of the hot electrons was measured using image plates (IPs) (Fujifilm BAS-2025). Image plates (IPs) were placed surrounding the target in a cylindrical geometry [Fig. 1(a)], in order to obtain fast electron angular distribution at the target front. To prevent the detection of low energy x-ray and electrons, IPs were covered with 110 $\mu$m-thick aluminium foil. The hot electron angular distribution was investigated as a function of laser intensity, which was varied by changing the energy of the incident laser pulse. The energy spectra of hot electrons were measured with a well calibrated magnet based electron spectrometer (ESM). Hot electrons spatially selected by an aperture pass through a region of uniform magnetic field (0.1 T) in vacuum. Magnetic field disperses the electrons according to their kinetic energies and the electrons are lodged on an IP-strip, which is read by an IP reader (FLA-7000 of FUJIFILM) to obtain the electron energy spectrum.\\

	\begin{figure}
		
		\includegraphics[width=1.0\columnwidth]{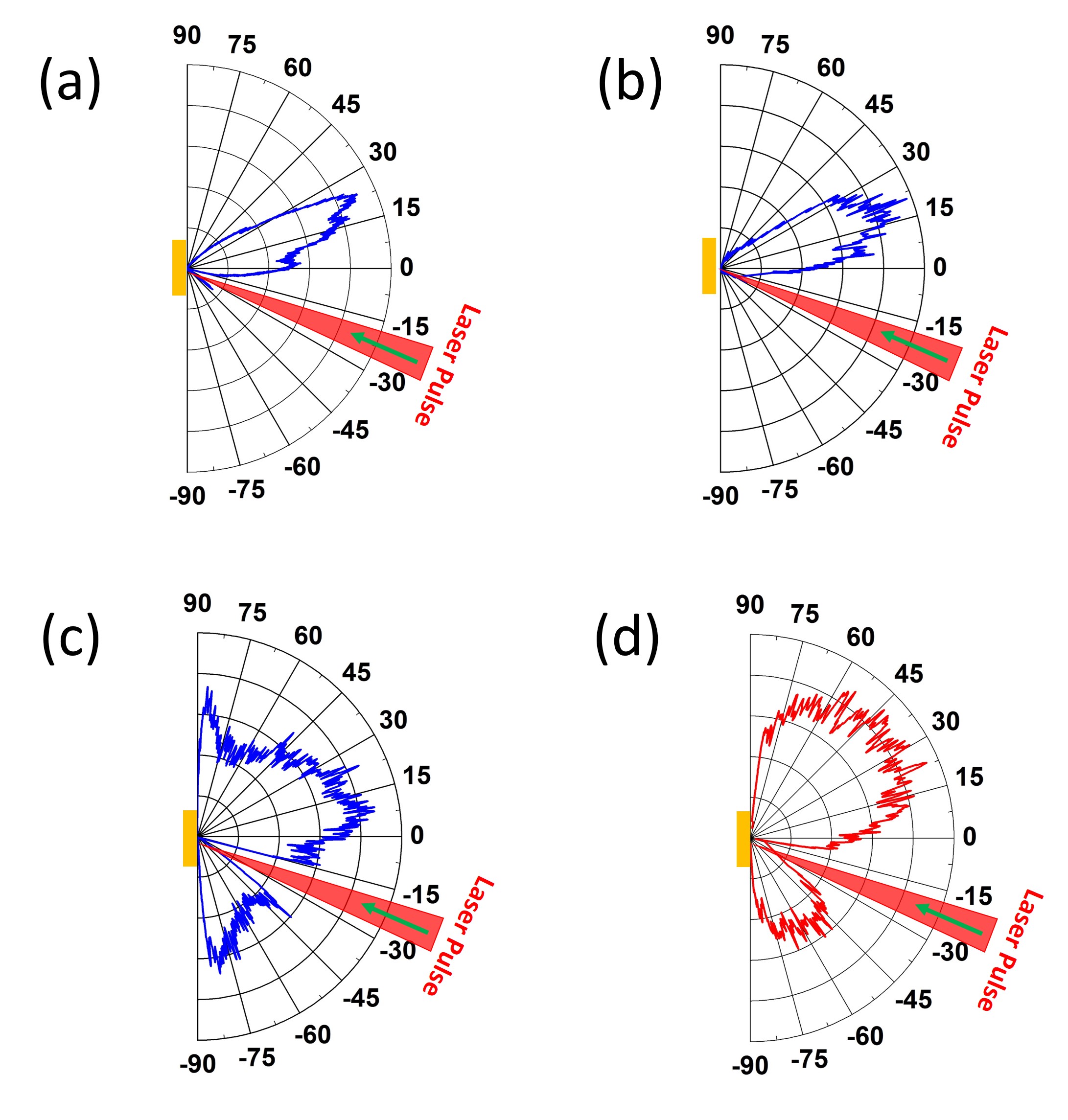}
		\caption{ Angular distributions of the hot electrons emitted from grating and mirror targets are shown for different laser intensities. Hot electron from grating target measured at (a) $I_L$ $\sim$ 3 $\times$ 10$^{18}$ W/cm$^2$ ; (b) $I_L$ $\sim$ 5 $\times$ 10$^{18}$ W/cm$^2$; (c) $I_L$ $\sim$ 1.2 $\times$ 10$^{19}$ W/cm$^2$ and (d) hot electrons from mirror target measured at $I_L$ $\sim$ 3 $\times$ 10$^{18}$ W/cm$^2$. Electrons arriving along the laser direction could not be measured.}
		\label{figure2}
		
	\end{figure}
	
	\section{Results}
	
	\begin{figure}
		
		\includegraphics[width=1.0\columnwidth]{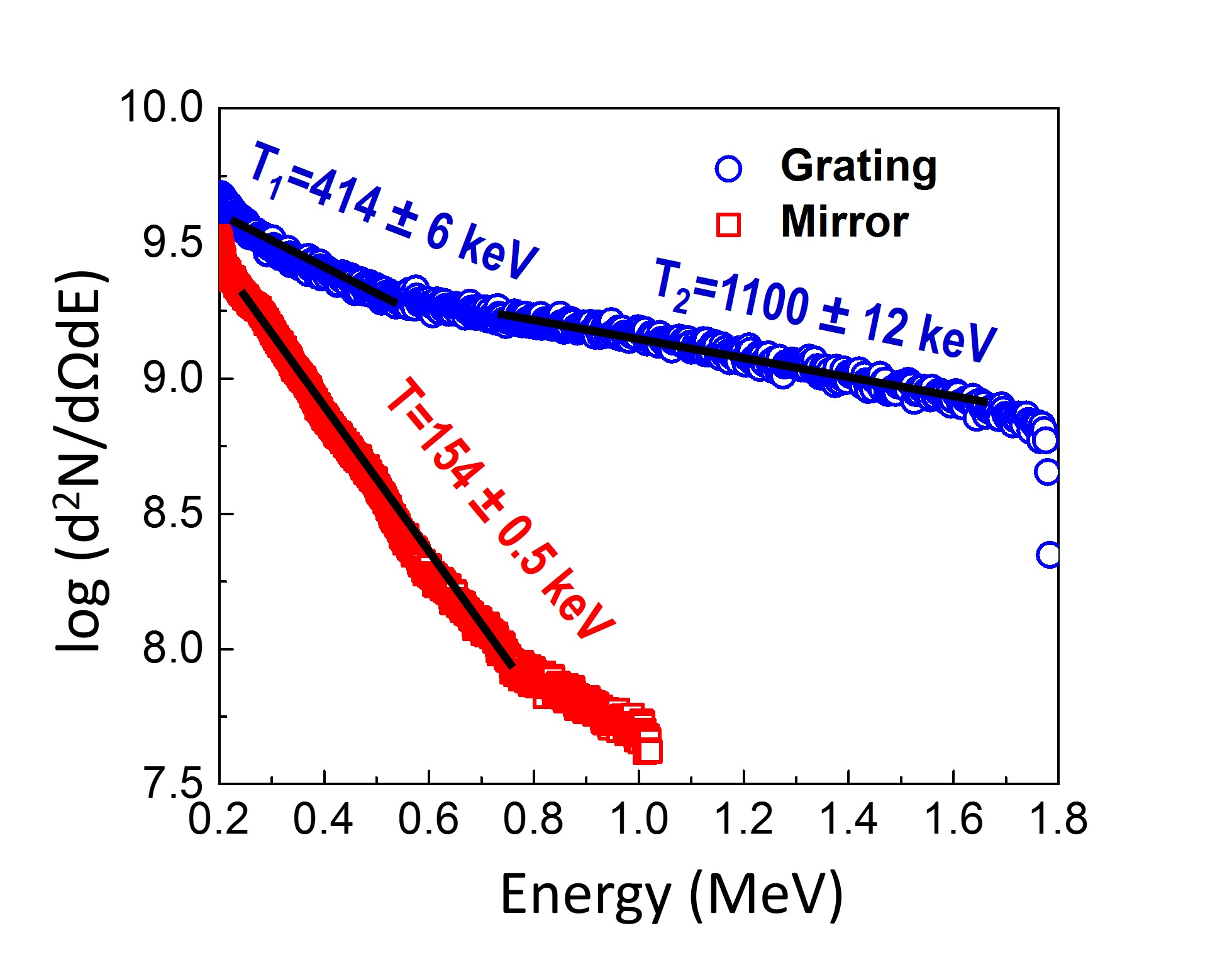}
		\caption{Energy spectra of hot electrons emitted from grating and mirror targets  measured at intensity $I_L$ $\sim$ 3 $\times$ 10$^{18}$ W/cm$^2$.}
		\label{figure3}
		
	\end{figure}

	The angular distributions of the hot electrons emitted from grating and plane mirror targets are presented in Fig. 2. We studied the effect of the laser intensity on the angular distributions of hot electrons generated from grating targets as shown in Fig. 2(a), 2(b) and 2(c). We see that collimated hot electrons are emitted nearly along the specular direction (zeroth order diffraction direction of the grating) at $I_L$ $\sim$ 3 $\times$ 10$^{18}$ W/cm$^2$. At a higher intensity $I_L$ $\sim$ 5 $\times$ 10$^{18}$ W/cm$^2$ the angular spread of the collimated electrons increases. At the highest intensity used ($I_L$ $\sim$ 1.2 $\times$ 10$^{19}$ W/cm$^2$) the emitted hot electrons spread over a large area. The angular distribution of the hot electrons generated from plane mirror target was also measured at intensity $I_L$ $\sim$ 3 $\times$ 10$^{18}$ W/cm$^2$ [Fig. 2(d)]. For this case, it is observed that hot electrons are emitted almost isotropically.

	Figure 3 represents the energy spectra of hot electrons from grating and mirror targets irradiated by ultrashort laser pulse at intensity $I_L$ $\sim$ 3 $\times$ 10$^{18}$ W/cm$^2$. Hot electron spectrum from grating target shows two temperature distribution with two characteristic temperatures 414 $\pm$ 6 keV and 1100 $\pm$ 12 keV. In a similar experimental configuration, electron spectrum for a plane mirror target shows a single temperature of only 154 $\pm$ 0.5 keV. The enhancement factor in integrated electron flux is 2.3 for the grating target with respect to plane target.\\

	\section{Simulations}
	A series of two-dimensional (2D) particle-in-cell (PIC) simulations were performed to understand the experimental observations by using the OSIRIS code. A p-polarized laser pulse with wavelength of $\lambda_0$=800 nm irradiates the targets at the incidence angle of 23$^{\circ}$. The peak laser intensity is $\sim$3$\times10^{18}$W/cm$^2$, which corresponds to the dimensionless amplitude of the electric field of $a_0$=1.186. The focal spot size (FWHM) and the pulse duration are 14 $\mu$m and 25 fs, respectively. Rectangular gratings are used in the simulation, with a groove period of 555 nm and the groove depth of 158 nm. These are almost consistent with the parameters in experiments. The plasma density of the grating and mirror targets in the simulations is 20$n_c$. The simulation box size is 35$\lambda_0$ $\times$ 80$\lambda_0$ with a cell size of $\lambda_0$/100 and 16 macro-particles per cell. In order to understand the mechanism of collimated hot electrons emission, we also studied the effects of the pre-plasma in front of the grating targets and the mirror target. The pre-plasma has a profile of $n(x)=n_{e}exp((x-20)/L)$ with the scale length $L$=0.1 and $L$=0.01 in simulations.
	
	\begin{figure}
		
		\includegraphics[width=1.0\columnwidth]{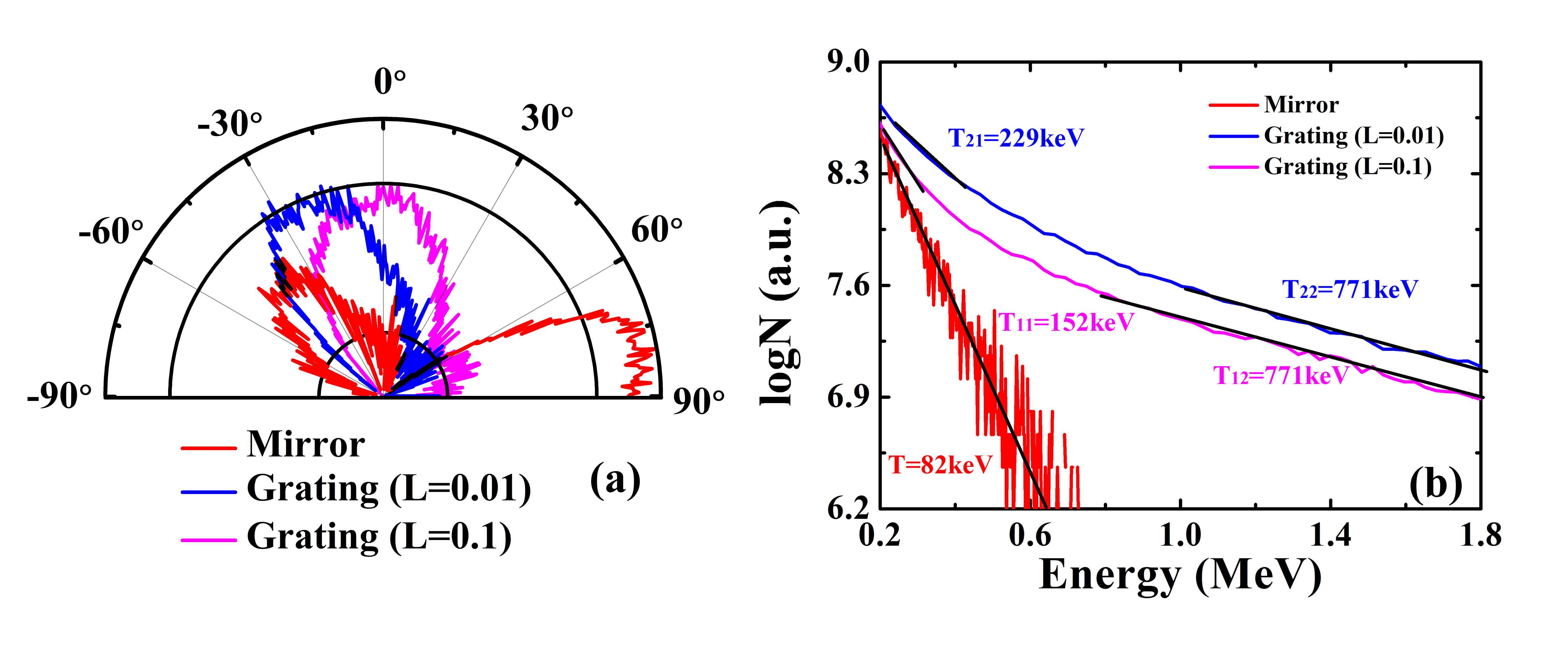}
		\caption{Simulated angular distributions (a) and energy spectra (b) of the hot electrons ($E_k\ge$50keV) recorded by the electron recorder located at the target front surface.}
		\label{figure4}
		
	\end{figure}

	\begin{figure}	
		\includegraphics[width=1.0\columnwidth]{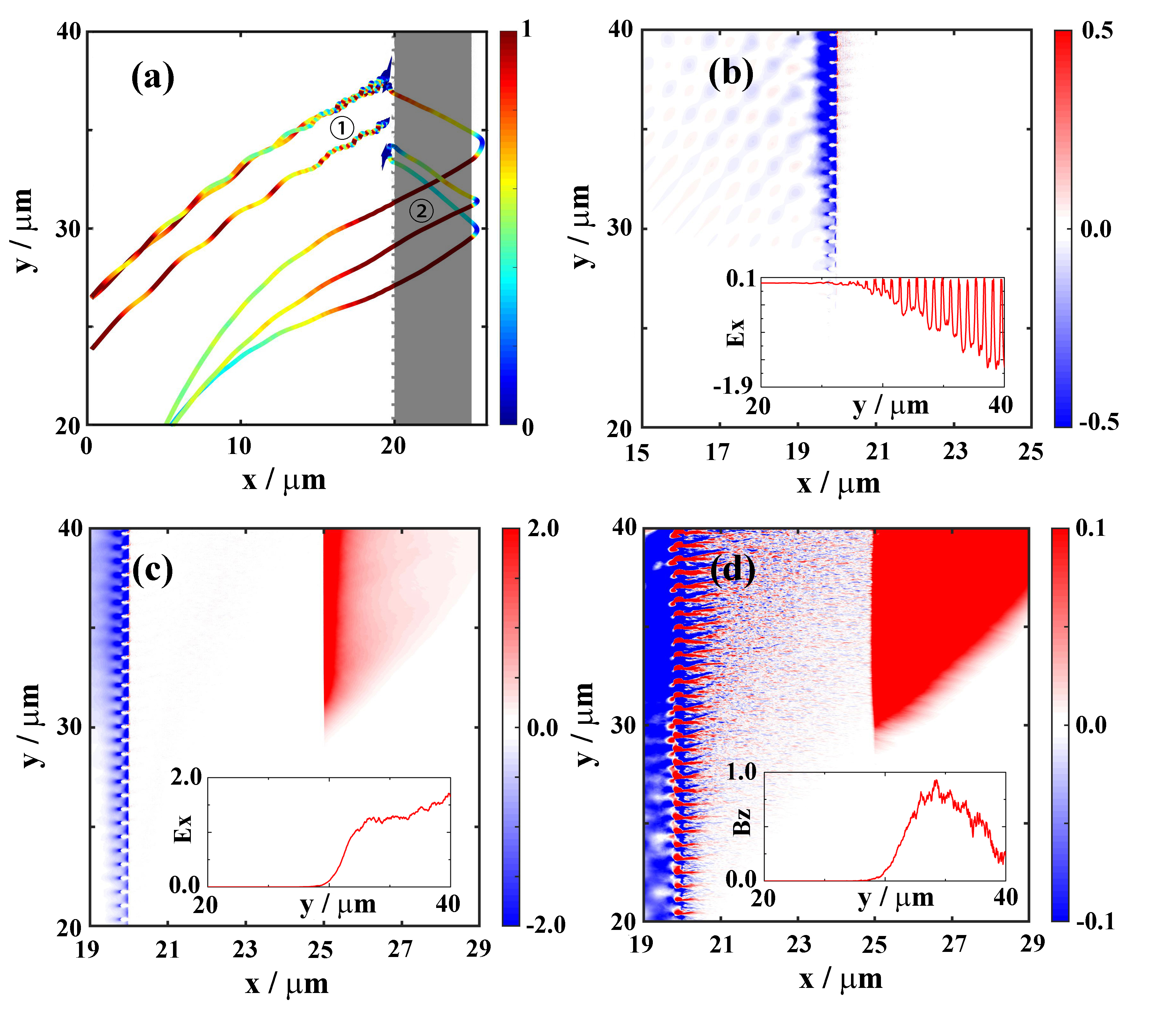}
		\caption{(a) Typical randomly selected electrons trajectories for grating target, and the color bar of trajectories represents the electron energy, 1 and 2 represent the two groups of the detected radiation electrons. The distribution of longitudinal electric field $E_x$ at $t$= 140 fs (b) and $t$= 173fs (c). The quasi-static magnetic field $B_z$ at $t$=173 fs (d). The inset in (b) shows the transverse distribution at $x$=19.9 $\mu$m, and the inset in (b) and (c) show the transverse distribution at $x$=25.5 $\mu$m. The fields are normalized by $m_ec{\omega}_p/e$}
		\label{figure5}
	\end{figure} 
	
	The angular distributions of the hot electrons emitted from mirror and grating targets are shown in Fig. 4(a). These electrons are recorded by the electron recorder located at the target front surface. Only the backward moving electrons ($p_x\le$0) with energy larger than 50 keV are recorded. The divergence angle of an electron is defined as $\theta=arctan(p_y/p_x$), where $p_y$ and $p_x$  are the momenta in $y$ and $x$ directions, respectively. One can see that a collimated hot electron bunch (due to the plasmon resonance) appears at the specular reflection direction with the scale length $L$=0.01 for the grating target. However, the emitted hot electrons spread out for the mirror target. At the scale length of $L$=0.1, the hot electrons are emitted mainly along the normal direction of the grating target due to the large pre-plasma length. This indicates that the pre-plasma length has an important effect on the angular distribution of the hot electrons. Figure 4(b) shows the energy spectra for different targets. The hot electron spectra of the grating target have two temperatures regardless of the pre-plasma, and for the mirror target, there is a single temperature. The simulation results are in good agreement with the experimental results.
	
	In order to understand the generation of the collimated hot electrons from the grating target, the electron dynamics are analyzed. Figure 5(a) shows typical (randomly selected) electrons trajectories and the color bar of the trajectories represents the electron energy. The detected electrons are mainly divided into two groups, the first group is the specular reflected electrons, and the second group is thermal electrons. Because of the surface plasma waves of the plasma grating, the laser carries the first group electrons along the laser reflection direction, but the number is small. For the second group electrons, due to the effect of grating structure, the electrons inside the front area of the protuberances are first stripped out from the grating by the laser electric field and accelerated by the longitudinal electric field $E_x$ on the front surface of the grating target. When these electrons move behind the target, they are reversely accelerated by the longitudinal electric field $E_x$ on the rear surface of the grating target, and are deflected by the quasi-static magnetic field $B_z$. The motion of the hot electrons is dominated by Lorentz equation, i.e.,  $d\theta/dL=eB_z/p$ \cite{MinOptExp2006}, and most of the electrons are emitted in a collimated beam, where $\theta$ is the emission angle, $L$ is the magnetic field length and $p$ is the electron momentum. To verify the hot electron emission, the longitudinal electric field at $t$= 140 fs [(Fig. 5(b)] (the electrons is located at the front of target) and at $t$=173 fs [Fig. 5(c)]  (the electrons is located at the rear of target) and quasi-static magnetic field at $t$=173 fs [Fig. 5(d)] are shown. The inset in Fig. 5(b) shows the transverse distribution at $x$=19.9 $\mu$m, and those in Figs. 5(c) and 5(d) show the transverse distribution at $x$=25.5 $\mu$m. The longitudinal electric and quasi-state magnetic fields are significantly enhanced due to higher energy absorption caused by the grating. The front surface longitudinal electric field accelerates the local electrons in protuberances, the rear surface electric field slows down the electrons and accelerates them in the reverse direction, and the magnetic field deflects the electrons. It is worth noting that the acceleration field and magnetic field perceived at different locations are different. 
	
	\begin{figure}
		
		\includegraphics[width=1.0\columnwidth]{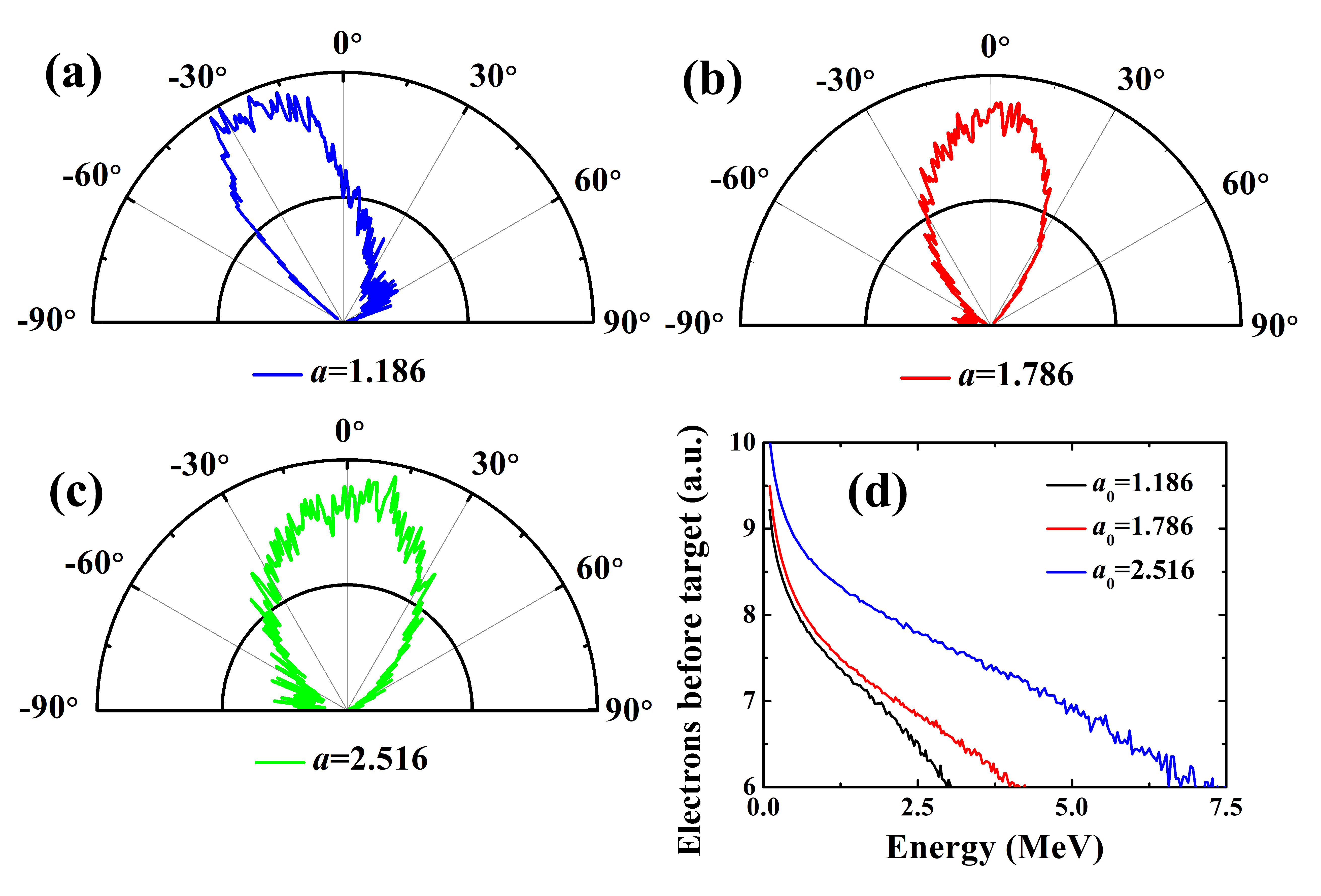}
		\caption{The simulated angular distribution (a), (b), (c) and energy spectra (d) of the hot electrons ($E_k\ge$50keV) for different laser intensity.}
		\label{figure6}
		
	\end{figure} 
	
	In addition, we also studied the effect of laser intensity on the angular distributions and energy spectra of the hot electrons, as shown in Fig. 6. We compare the emitted hot electrons for three different laser intensities, $I_L\sim3\times10^{18}$ W/cm$^2$ ($a_0$=1.186), $I_L\sim6.8\times10^{18}$ W/cm$^2$ ($a_0$=1.786), $I_L\sim13.5\times10^{18}$ W/cm$^2$ ($a_0$=2.516). In these simulations, the plasma scale length is fixed at $L$=0.01, and other parameters are the same as before. The simulation results show that the angular distributions of the electrons vary greatly with the laser intensity. Only at low laser intensity, the hot electrons radiate along the specular reflection direction. At high laser intensity, the hot electrons emitted along the normal direction of the grating target due to the destruction of the grating structure, which is consistent with the case of large plasma density scale length in Fig. 4(a). The electron energy spectra for different laser intensities are shown in Fig. 6(d). One can see that the number, maximum electron energy and temperature of the emitted hot electrons increase with the laser intensity. The maximum temperature of the hot electrons achieved is 870 keV for an intensity of $a_0$=1.786 and 1170 keV for $a_0$=2.516.  \\

	\section{Conclusion}
	
	In summary, we investigated fast electron generation from sub-wavelength grating targets irradiated by relativistic laser pulses. Emission of highly collimated energetic fast electrons from the grating target was observed, along the near specular direction (zeroth order diffraction) of the grating target. Fast electron emission as a function laser intensity were measured. Experimental observations reveal efficient generation of higher temperature and more collimated electrons from grating target than those from polished mirror target. Numerical simulations well reproduced experimental observations and indicates efficient laser energy coupling to the periodic grating target and electron beam steering by the resulting electric and magnetic field at the grating surface. The intense laser-driven collimated relativistic electrons from grating will find many potential applications in basic, medical and industrial sciences.\\

	\section*{ACKNOWLEDGMENTS}
	
	GRK acknowledges partial support from a J. C. Bose Fellowship grant (JBR/2020/000039) from the Science and Engineering Research Board, Government of India. MC acknowledges the support by NSFC (11991074, 11774227) of China, Natural Science Foundation of Shandong Province (No. ZR2019ZD44) and the Strategic Priority Research Program of Chinese Academy of Sciences (grant no. XDA25000000). The authors would like to acknowledge the OSIRIS Consortium, consisting of UCLA and IST (Lisbon, Portugal) for the use of OSIRIS and the visXD framework. The authors also thank Sudipta L. Roy for his help in the experiments.

	\section*{DATA AVAILABILITY}
	The data that support the findings of this study are available within the article.

\end{document}